\RequirePackage{etex}
\documentclass[10pt, twoside]{predoc2}

\usepackage[RGB,table,dvipsnames]{xcolor}
\usepackage{graphicx}
\usepackage{tabularx}
\usepackage{booktabs}
\usepackage{subcaption}
\usepackage{hyperref}
\usepackage{cleveref}
\usepackage{tikz}
\usepackage{tikz-network}
\usepackage{newfloat}
\usepackage{orcidlink}

\usepackage{geometry}
\usetikzlibrary{matrix,positioning,quotes,fit,3d,arrows.meta,shapes,calc,shapes.geometric}

\definecolor{custom1}{RGB}{1, 115, 178}
\definecolor{custom2}{RGB}{222, 143, 5}
\definecolor{custom3}{RGB}{2, 158, 115}
\definecolor{custom4}{RGB}{213, 94, 0}
\definecolor{custom5}{RGB}{204, 120, 188}
\definecolor{custom6}{RGB}{202, 145, 97}
\definecolor{custom7}{RGB}{251, 175, 228}
\definecolor{custom8}{RGB}{148, 148, 148}
\definecolor{custom9}{RGB}{236, 225, 51}
\definecolor{custom10}{RGB}{86, 180, 233}

\fancypagestyle{firstpage}{%
  \fancyhf{}

  \fancyfoot[C]{\footnotesize Preprint. \itshape
    Correspondence: Dominic Gr{\"u}n (\href{mailto:dominic.gruen@uni-wuerzburg.de}{dominic.gruen@uni-wuerzburg.de})
    and Ingo Scholtes (\href{mailto:ingo.scholtes@uni-wuerzburg.de}{ingo.scholtes@uni-wuerzburg.de})}
}

\DeclareFloatingEnvironment[
  fileext=loe,
  name=Extended Data Fig.,
  placement=tbp,
]{edfigure}

\crefname{edfigure}{Extended Data Fig.}{Extended Data Figs.}
\Crefname{edfigure}{Extended Data Fig.}{Extended Data Figs.}

\title{Single-Snapshot Inference of Network Couplings from Universal Dynamics at Relative Equilibrium}

\author[1,2]{Moritz Lampert \orcidlink{0009-0007-6692-5646}}
\author[1,3,$\ast$]{Dominic Gr{\"u}n \orcidlink{0000-0002-3364-5898}}
\author[1,2,$\ast$]{Ingo Scholtes \orcidlink{0000-0003-2253-0216}}
\affil[1]{CAIDAS - Center for Artificial Intelligence and Data Science, W{\"u}rzburg, Germany.}
\affil[2]{Chair of Machine Learning for Complex Networks, Julius-Maximilians-Universit{\"a}t, W{\"u}rzburg, Germany.}
\affil[3]{Würzburg Institute of Systems Immunology, Julius‐Maximilians-Universit{\"a}t W{\"u}rzburg, W{\"u}rzburg, Germany.}
\affil[$\ast$]{Co-last and co-senior authors}
\date{}

\begin{document}
\maketitle
\thispagestyle{firstpage}
\begin{abstract}
    Many real-world systems can be modelled as complex networks whose collective behaviour is governed by hidden interactions between nodes.
    Existing methods for inferring these interactions typically require controlled perturbations, time-resolved observations or multiple independent snapshots, all of which are often unavailable in practice.
    Here we show that class-based coupling strengths can be inferred from a single snapshot of node states when the system is observed close to a relative equilibrium.
    In this regime, all nodes share a common velocity, which can be absorbed into an effective class bias, transforming the inverse problem into a homogeneous linear system.
    The coefficients of this linear system are determined entirely by the observed local neighbourhoods and their coupling mechanism, enabling the application to arbitrary known coupling functions.
    We validate the approach on three different linear and nonlinear dynamical systems, recovering relative class-based couplings and, in special cases, absolute couplings.
    These results show that spatial heterogeneity can substitute for temporal sampling, enabling single-snapshot inference of hidden coupling strengths in networked dynamical systems.
\end{abstract}
%

Interactions between nodes shape the dynamics of many physical, biological and social systems.
Node states such as neural activity, gene expression, oscillator phases, and opinions can often be measured at the level of individual nodes, but the couplings that determine how nodes influence one another are usually hidden.
Inferring these couplings from observed states is therefore a central inverse problem in network dynamics.

Because the coupling strengths govern how node states evolve, their inference generally requires information about the system's temporal evolution.
Thus, existing approaches depend on observing the dynamics~\cite{RevealingNetworksDynamics2014timme,InverseStatisticalProblems2017nguyen,DataBasedIdentification2016wang,ConnectingDotsIdentifying2019mateos}. 
One line of work infers couplings from the system's response to controlled perturbations~\cite{RevealingNetworkConnectivity2007timme} or natural variations in its state~\cite{InferringNetworkTopology2011shandilya}.
A related class of methods reconstructs interactions from steady-state data rather than full trajectories, but still requires systematically perturbing the system across many experimental conditions, as in gene-regulatory network reconstruction~\cite{InterpretationExtrapolationPerturbation2026dimitrov}.
Others reconstruct topology or coupling weights from high-resolution time series, using model-based reconstruction~\cite{EstimatingTopologyNetworks2006yu}, model-free inference~\cite{ModelfreeInferenceDirect2017casadiego}, sparse identification~\cite{DiscoveringGoverningEquations2016brunton,InferringBiologicalNetworks2016mangan,HypergraphReconstructionDynamics2025delabays}, compressed sensing~\cite{TimeseriesbasedPredictionComplex2011wang,RobustReconstructionComplex2015han}, or Neural Relational Inference~\cite{NeuralRelationalInference2018kipf}.
Even methods that do not rely on high-resolution time series data still require repeated observations~\cite{NetworkStructureRich2018newman} of graph signals~\cite{NetworkTopologyInference2017segarra,NetworkInferenceConsensus2020zhu} or discrete-time dynamics~\cite{NetworkReconstructionBased2011wang}, temporally disordered observations collected across experiments~\cite{RevealingPhysicalInteraction2017nitzan}, or multiple independent snapshots~\cite{UnsupervisedRelationalInference2023grossmann}.
A single snapshot captures only the node states, not their rate of change, making the underlying couplings hard to infer.
While specialised approaches, such as RNA velocity, can recover this rate of change from a single snapshot for individual genes ~\cite{RNAVelocitySingle2018lamanno}, such approaches typically treat individual genes as independent nodes, ignoring regulatory couplings.
A recent method recovers a few global parameters of a known dynamical law from a single snapshot~\cite{EfficientParameterInference2024ding}, but no node-to-node couplings.

Here, we show that a single snapshot can suffice to recover these couplings, provided it is taken when the system is close to a relative equilibrium.
The inference is possible in this case because of two properties:
First, all nodes share a common velocity.
We do not need to measure how the node states evolve because the uniform constant is absorbed into the inferred parameters.
Since absorbing the constant removes the dependence on a time derivative, we do not require a second observation.
Second, because the coupling strengths govern the relative-equilibrium configuration, a single snapshot of the node states already constrains them -- similar to the way the resting state of a spring network constrains the stiffnesses connecting its nodes.
A single node imposes only one such constraint, but a network comprises many nodes in distinct local environments.
Our approach draws the variation it needs from these differences across space, where time-series methods draw the same variation from successive points in time.

We establish this for networks whose nodes belong to known classes with unknown, class-based coupling strengths and recover the couplings analytically. An important and timely application of this scenario applies to the reconstruction of spatial molecular interaction networks between cells in tissues. Here, cells belong to different types corresponding to distinct classes, and each cell can respond to molecular signals from neighbouring cells by modulating gene programs. Inferring class-specific couplings between these gene programs in neighbouring cells provides important biological information on tissue biology
\cite{NiCoIdentifiesExtrinsic2024agrawal}.
Building on network-based neighbourhood regression~\cite{NetworkbasedNeighborhoodRegression2025zhen}, which expresses each node's state as a linear combination of its neighbours' states, we cast the recovery of class-based coupling strengths from a single snapshot as a linear regression problem.
This formulation is universal, which means it applies to arbitrary known coupling mechanisms.
As long as the functional form of the coupling is known, this inverse problem can be solved for both linear and nonlinear dynamical systems.

\begin{figure}
    \centering
    \input{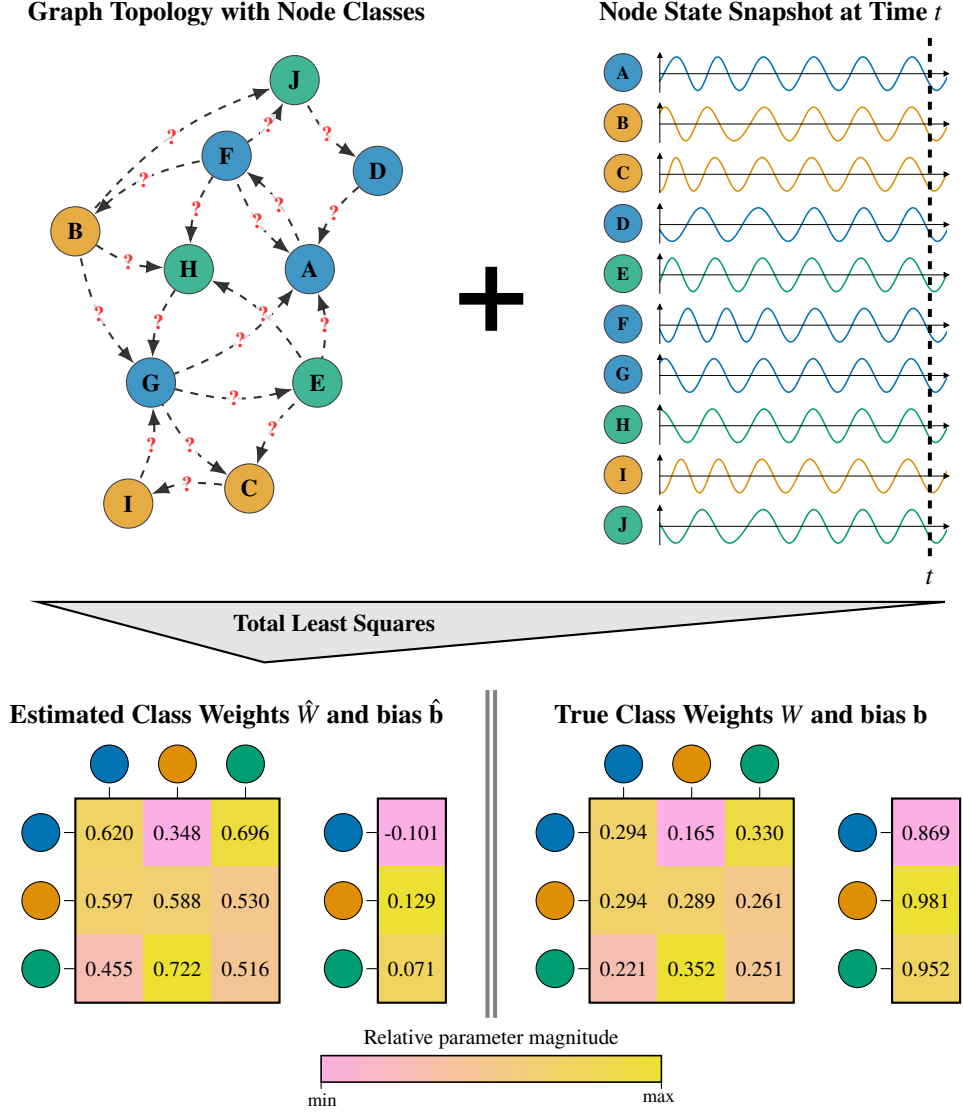}%
    \caption{\textbf{Overview of the proposed parameter inference method.} Given a graph where the topology and node class assignments are known (top left), we can infer hidden class-based coupling weights and class-intrinsic biases (bottom left) using only a single instantaneous snapshot of the node states at time $t$ (top right). The displayed matrices and state trajectories represent an actual simulation of Kuramoto oscillator dynamics on a 10-node network. For the inference to succeed, the system must exhibit spatially uniform velocity at the measurement time $t$ -- which, in this Kuramoto example, corresponds to a phase-locked state. Because our method formulates the inference as a homogeneous system, the true parameters (bottom right) can be recovered only up to an unknown scale and shift. Consequently, the absolute numerical values of the estimated parameters $\hat{W}$ and $\hat{\mathbf{b}}$ naturally differ from the ground truth $W$ and $\mathbf{b}$. However, as demonstrated by the matching colour gradients in the heatmaps, the relative coupling strengths between classes are recovered near-perfectly via Total Least Squares.}
    \label{fig:overview}
\end{figure}

\paragraph{Learning Relative Class-based Coupling Strengths}
We consider a dynamical system on a network $G = (V, E)$ without self-loops.
The network consists of nodes $v \in V$, where each node belongs to one specific class $z_v \in \{1, \dots, c\}$.
The temporal evolution of the state variable $x_v$ at node $v$ is governed by a class-based intrinsic bias $b_{z_v}$ and the coupling with the nodes in its neighbourhood $N(v)$.

For a given edge $(v, u)\in E$, we assume that the coupling strength $W_{z_v,z_u}$ and the bias $b_{z_v}$ are determined by the class assignments of the connected nodes.
We define the continuous-time dynamics of the system as
\begin{align}\label{equ:general_dynamical_system}
    \dot{x}_v = b_{z_v} + \sum_{u \in N(v)} W_{z_v,z_u} f(x_v,x_u)
\end{align}
where $f(x_v,x_u)$ defines the coupling mechanism along the edges. 
With this, we obtain a universal formulation of network dynamics similar to \citet{UniversalityNetworkDynamics2013barzel}, with the difference that we consider class-based couplings.

By transforming this system to a homogeneous system of linear algebraic equations whose right-hand side is zero, we show that we can reconstruct the relative class-based coupling strengths $W^*_{z_v,z_u}$ and the relative class-based effective biases $\tilde{b}_{z_v}$ of a dynamical system (see Methods for details).
For this, we only need a single snapshot of all node states $x_v$ at one arbitrary point in time if the following conditions are satisfied: 
(i) The dynamical system follows the functional form specified in \Cref{equ:general_dynamical_system} with class-specific biases $b_z$ that are sufficiently heterogeneous, and the functional form $f$ of the coupling mechanism is known a priori.
(ii) The rank of the design matrix $M_z$ constructed from the class-based local neighbourhood (see Methods) has full column rank $c$.
(iii) The snapshot is taken at a relative equilibrium, that is, an instant of uniform node velocity $\dot{\mathbf{x}} = k\mathbf{1}$.
In particular, the node states do not need to be constant.
As long as they all evolve at a common rate $k$, we can treat the configuration as stationary once this shared collective component is accounted for.
These conditions enable the inference of class-based coupling strengths analytically based on the total least squares method, both for linear and nonlinear coupling mechanisms $f$.
Crucially, the single snapshot evaluates $f$ on the observed states and thereby freezes the nonlinearity into fixed coefficients of the design matrix, rendering the inference linear in the unknown $W$ and $b$ for any known coupling mechanism $f$.
See \Cref{fig:overview} for a visual overview of our approach and the Methods section for full mathematical details.

To validate the efficacy of our regression approach, we conduct controlled experiments with simulated dynamical systems on networks in the following.
We use three different types of linear and nonlinear dynamical systems to demonstrate the versatility of our method.

\paragraph{Linear Consensus Dynamics} We begin our validation with a linear dynamical system governed by a diffusive coupling mechanism $f(x_v, x_u) = x_u - x_v$.
The evolution of the node states is then given by
\begin{align}\label{equ:consensus_dynamical_system}
    \dot{x}_v = b_{z_v} + \sum_{u \in N(v)} W_{z_v,z_u} (x_u - x_v).
\end{align}
This type of Laplacian dynamics serves as a canonical starting point for our validation, as its well-understood spectral properties~\cite{chung1997spectral} — in particular, the role of algebraic connectivity~\cite{fiedler1973algebraic} in convergence behaviour — allow for a systematic characterisation of when our inference conditions are met.

\begin{figure}[tb]
    \input{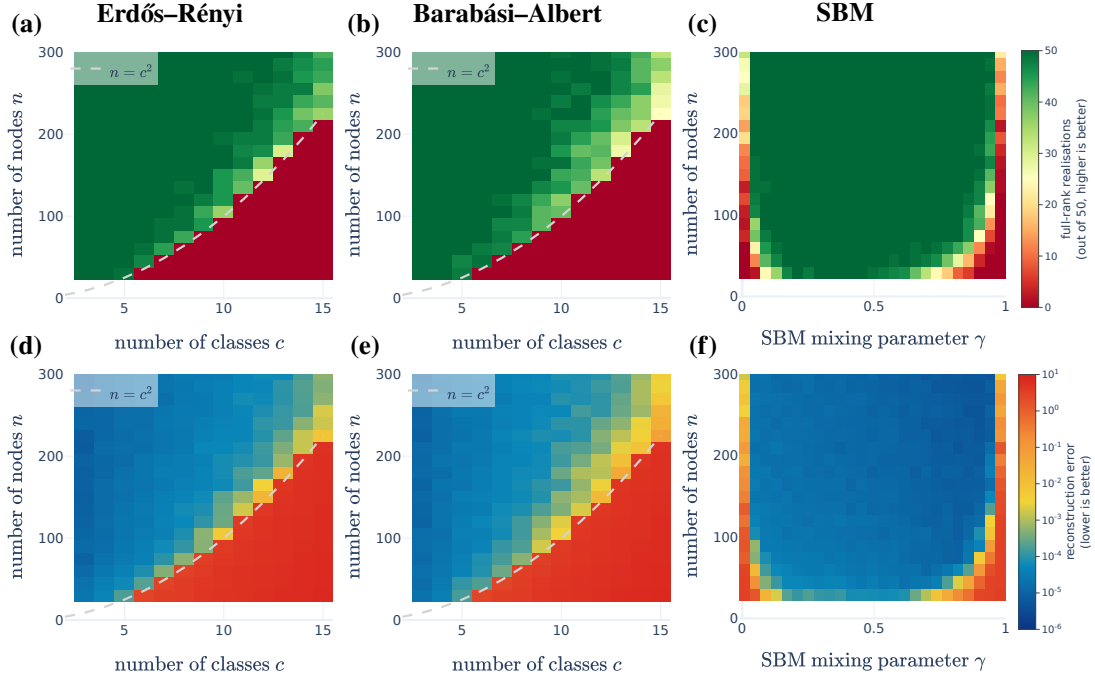}
    \caption{\textbf{Rank sufficiency and reconstruction error.} \textbf{(a--c)} Heatmaps displaying the number of times the rank condition $\#\mathrm{r}(M_z)=c$ is satisfied out of 50 random initialisations for varying network sizes and topologies. \textbf{(d--f)} Corresponding heatmaps showing the mean normalised reconstruction error $||\hat{\tilde{W}}^* - \tilde{W}^*||_F$ on a logarithmic colour scale. The random realisations are drawn from: \textbf{(a, d)} Erd\H{o}s-R\'{e}nyi random graphs with edge probability $p=\frac{8}{n}$; \textbf{(b, e)} Barab{\'a}si-Albert random graphs with $m=8$ new edges per node; and \textbf{(c, f)} SBMs with fixed $c=5$ classes and a varying mixing parameter $\gamma$. The edge probability is $p_{z,\tilde{z}}=8\cdot \frac{c}{n} \cdot \frac{1-\gamma}{c-1}$ if $z\not=\tilde{z}$, and $p_{z,\tilde{z}}=8\cdot \frac{c}{n} \cdot \gamma$ otherwise (with all probabilities capped at 1). The grey dashed line in the Erd\H{o}s-R\'{e}nyi and Barab{\'a}si-Albert plots denotes the theoretical minimum network size ($n=c^2$) required for every class to contain $c$ nodes, so that each class-wise design matrix $M_z$ attains full column rank $c$. Nodes are partitioned into $c$ near-balanced blocks (each of $\lfloor n/c\rfloor$ nodes, with the last class absorbing any remainder); weights and biases are drawn from normal distributions: $W_{z_v,z_u}\sim \mathcal{N}(0.2,0.1)$ and $b_{z_v}\sim \mathcal{N}(0.5,0.25)$. Simulations are run on the linear consensus system (\Cref{equ:consensus_dynamical_system}) until a relative equilibrium of uniform node velocity is reached, quantified by the standard deviation of the node velocities ($\mathrm{std}(\dot{\mathbf{x}}) < 10^{-6}$), using an explicit adaptive Runge-Kutta method of order 5(4). $\hat{\tilde{W}}^*$ and $\tilde{W}^*$ are the normalised estimated and ground truth augmented parameter matrices $\hat{\tilde{W}}$ and $\tilde{W}$ (see Methods).}
    \label{fig:rank_vs_L2}
\end{figure}

We start by verifying the rank condition of the design matrix (Condition (ii)) and simulate this dynamical system on random graphs with different sizes and topologies.
\Cref{fig:rank_vs_L2} shows the influence of the network size $n$ on the rank sufficiency and the reconstruction error of our approach.
Rank sufficiency captures whether each class-wise design matrix $M_z$ has full column rank, which we use as a non-degeneracy criterion for the total-least-squares problem.
Specifically, the colour gradient in the top heatmaps displays the number of realisations (out of 50 random initialisations) that fulfil the rank condition for specific combinations of node and class counts.
The reconstruction error of $\hat{\tilde{W}}^*$ (see Methods) quantifies the level of accuracy to which we can reconstruct the class-based relative coupling strengths and is shown using a log-scale colour gradient in the heatmaps of the bottom row.
In the two leftmost columns, we consider sparse Erd\H{o}s-R\'{e}nyi random graphs \cite{RandomGraphs1959erdos} with a mean degree of eight as well as Barab{\'a}si-Albert random graphs \cite{EmergenceScalingRandom1999barabasi} in which each newly added node attaches with $m=8$ edges.
The grey dashed line represents the theoretical minimum number of nodes to achieve sufficient rank.
Specifically, $c^2$ nodes are required for at least $c$ non-zero rows in the design matrix for each class.
We observe that below the theoretical minimum network size ($n=c^2$), a unique solution is never obtained.
At or just above the theoretical minimum, the rank condition is satisfied in almost all realisations, irrespective of whether the topology exhibits a uniform (Erd\H{o}s-R\'{e}nyi random graphs in the left column) or power-law (Barab{\'a}si-Albert random graphs in the centre column) degree distributions.
In the rightmost column, we additionally consider random graphs generated by a Stochastic Block Model (SBM) with $c=5$ classes \cite{StochasticBlockmodelsFirst1983holland}, where we vary a mixing parameter $\gamma$ that governs the probability of inter- vs. intra-class edges.
As the mixing parameter $\gamma$ approaches 0, the graphs contain only inter-class edges and become multipartite; as it approaches 1, they contain only intra-class edges and fragment into disconnected components.
Both extremes lead to topological degeneracies that violate the rank sufficiency condition, as shown by the small number of realisations (red colour) where the rank sufficiency is met in \Cref{fig:rank_vs_L2}c.

The average reconstruction error correlates with the rank sufficiency, decreasing in parameter regimes where the rank sufficiency condition is met.
Furthermore, we note that the minimum reconstruction error is smaller for Barab{\'a}si-Albert random graphs, which reach the stationary drift state faster (at $t=23.73$ on average) compared to Erd\H{o}s-R\'{e}nyi random graphs (average $t=69.92$).
This suggests that the accuracy of our inference method is influenced by the speed at which the network topology facilitates consensus.

\begin{figure}
    \centering
    \input{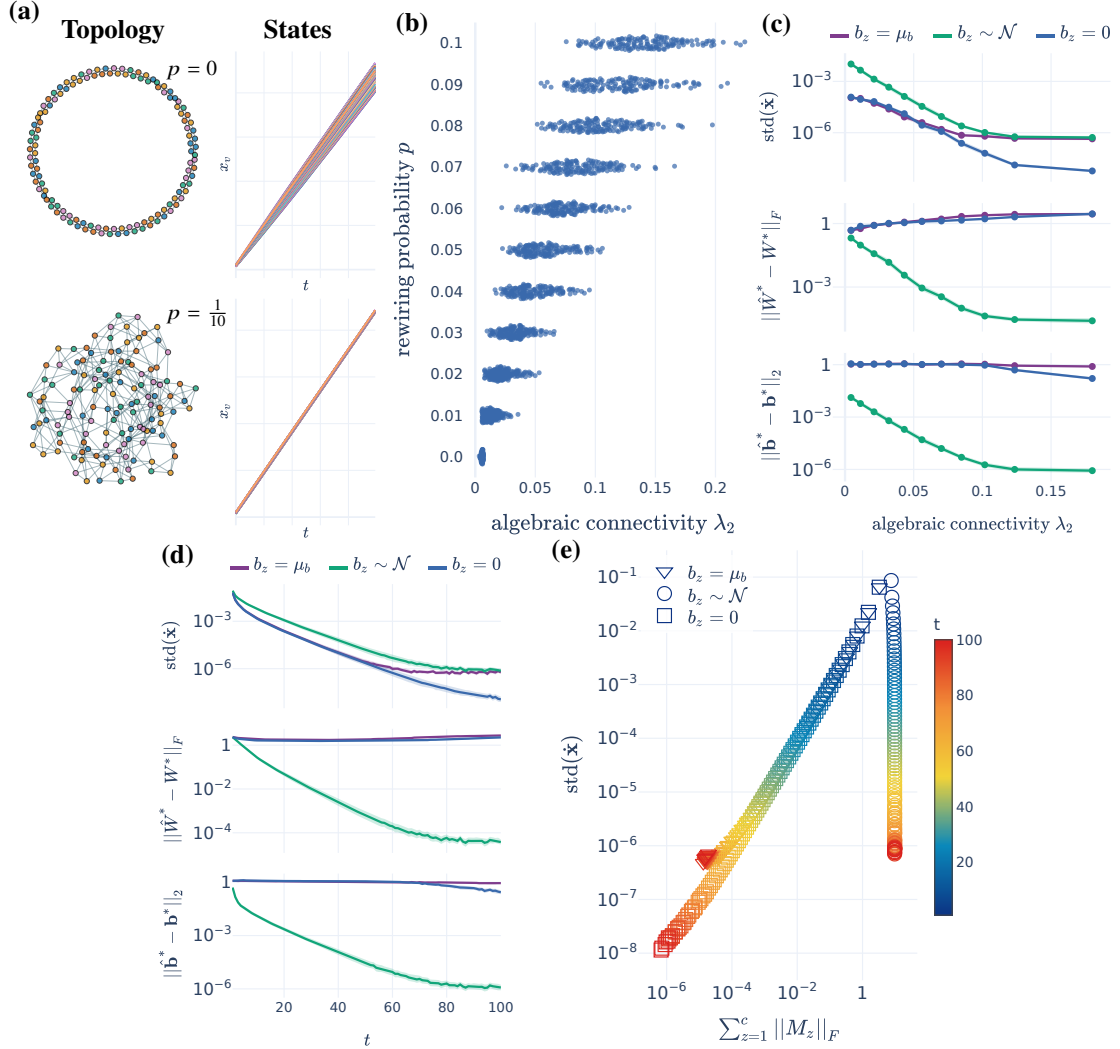}%
    \caption{\textbf{Experiments on linear consensus dynamics.} \textbf{(a)} Topologies and corresponding node state trajectories for Watts-Strogatz random graphs with no ($p=0$) and high ($p=0.1$) rewiring probability, illustrating how an increased number of ``shortcuts'' accelerates consensus. \textbf{(b)} The algebraic connectivity $\lambda_2$ (the second smallest eigenvalue of the weighted Laplacian) as a function of the rewiring probability $p$. Results are based on 50 Watts-Strogatz graphs with $n=125$ nodes connecting to their two nearest neighbours and $c=5$ uniformly distributed classes. \textbf{(c)} Standard deviation of the node state velocity $\operatorname{std}(\dot{\mathbf{x}})$ and the reconstruction error of the normalised weights and biases evaluated at $t=100$ across different topologies. Systems with intrinsic bias are compared against systems with zero and a constant bias. Measurements are binned into 11 equal-sized quantiles, displaying the mean and $95\%$ confidence intervals. \textbf{(d)} The same metrics tracked over time $t$ for Watts-Strogatz graphs with a fixed rewiring probability of $p=0.1$, measured at integer time steps. \textbf{(e)} A state-space projection of the temporal evolution from \textbf{(d)}, mapping the deviation from uniform velocity $\operatorname{std}(\dot{\mathbf{x}})$ against the preserved structural signal in the design matrices $\sum_{z=1}^{c}\|M_z\|_F$. The colour gradient represents the progression of time $t$. Each trajectory point is averaged across 50 realisations. For all experiments, simulations are done according to the experiments in \Cref{fig:rank_vs_L2}. $W$ and $\mathbf{b}$ are sampled from a normal distribution with $\mu_W = \mu_{\mathbf{b}}=0.5$ and $\sigma_W=0.1$ and $\sigma_{\mathbf{b}}=0.25$. The asterisk $*$ marks matrices and vectors as normalised (see Methods).}
    \label{fig:equilibrium_diffusion}
\end{figure}

The convergence rate of linear consensus dynamics is governed by the algebraic connectivity $\lambda_2$, i.e. the second smallest eigenvalue of the Laplacian matrix, which captures how ``well-connected'' the graph is~\cite{fiedler1973algebraic}.
In Watts-Strogatz random graphs, we can control the algebraic connectivity \cite{CollectiveDynamicsSmallworld1998watts} using the rewiring probability $p$ (see \Cref{fig:equilibrium_diffusion}a for an illustration).
\Cref{fig:equilibrium_diffusion}b verifies that a larger rewiring probability $p$ applied to an initial ring lattice results in larger algebraic connectivities $\lambda_2$, which in turn leads to faster consensus.

To assess how close the system is to relative equilibrium, in the top panel of \Cref{fig:equilibrium_diffusion}c, we investigate the standard deviation of the node velocities $\operatorname{std}(\dot{\mathbf{x}})$ (see Methods for details) at a fixed snapshot $t=100$ for Watts-Strogatz random graphs with different algebraic connectivities $\lambda_2$.
We sample coupling weights and heterogeneous class-based biases from a normal distribution. 
To validate the necessity of class-based biases $b_z$, we evaluate two control scenarios: a system with no bias ($b_z=0$ for all classes $z$) and a system with uniform bias ($b_z=\mu_{\mathbf{b}}$ for all $z$ that matches the mean of normally distributed biases $b_z \sim \mathcal{N}$).
As expected, networks with larger $\lambda_2$ are closer to a relative equilibrium, as evidenced by a decrease in the deviation from uniform velocity $\operatorname{std}(\dot{\mathbf{x}})$.

In the lower panels, we assess the normalised average reconstruction error for the class-based relative coupling strengths ($\|\hat{W}^*-W^*\|_F$) and the class-based relative effective biases ($\|\hat{\mathbf{b}}^*-\mathbf{b}^*\|_2$) for Watts-Strogatz graphs with different $\lambda_2$.
We find that the deviation from uniform velocity drops as the algebraic connectivity increases.
In line with our expectations, this velocity uniformity -- which satisfies the theoretical assumption of our homogeneous system formulation -- correlates with the reconstruction errors for class-based coupling strengths and heterogeneous class biases that approach zero as $\lambda_2$ increases.
For uniform class biases $b_z=\mu_b$ as well as zero biases $b_z=0$ we observe high reconstruction errors regardless of algebraic connectivity.
This demonstrates that achieving uniform velocity is a necessary but not sufficient condition. 
Our method additionally requires an intrinsic bias that is unique to each class to succeed.

In \Cref{fig:equilibrium_diffusion}d, we examine the temporal evolution of the quantities from \Cref{fig:equilibrium_diffusion}c for different times $t$ to pinpoint why our approach fails with uniform or zero bias.
The standard deviation (top panel) vanishes over time in all systems regardless of bias, indicating that they all reach a relative equilibrium.
Yet only for the system with heterogeneous class-based biases do the reconstruction errors decrease (bottom panels).
\Cref{fig:equilibrium_diffusion}e visualises the reason why the reconstruction fails without a class-based bias. 
For this, we project the temporal evolution into a state space defined by the deviation from uniform velocity (y-axis) and the sum of the Frobenius norms of the design matrices of all classes (x-axis).
This norm indicates how much information we can obtain about class-based couplings from a given snapshot $t$.
We observe that the trajectories of systems with uniform or no bias collapse to the origin:
they satisfy the uniform velocity requirement, but thereby destroy the information needed to infer class-based couplings.
Without class-based biases that act as heterogeneous external driving forces, all pairwise state differences vanish.
This effectively destroys the information within the design matrix ($M_z \to \mathbf{0}$).
In contrast, the heterogeneous class-intrinsic bias forces the dynamical system into a state in which a uniform velocity is achieved while preserving information about class-based couplings.

\paragraph{Nonlinear Kuramoto Dynamics}
The interactions in many real-world networks are inherently nonlinear. Prominent examples include pacemaker cells in the heart~\cite{MechanismsSinoatrialPacemaker1987michaels}, circadian cells in the brain~\cite{CellularConstructionCircadian1997liu}, microwave oscillators~\cite{QuasiopticalPowerCombining1991york}, and superconducting Josephson junctions~\cite{FrequencyLockingJosephson1998wiesenfeld}, all of which exhibit synchronisation phenomena that a linear coupling cannot capture. 
One of the benefits of our approach is that it makes no assumptions about the coupling mechanism $f(x_v, x_u)$, meaning we can seamlessly apply it to such nonlinear dynamical systems. 
To demonstrate this, we evaluate our inference method on the Kuramoto model of coupled phase oscillators~\cite{SelfentrainmentPopulationCoupled1975kuramoto,KuramotoModelSimple2005acebron}, which serves as a canonical model for synchronisation phenomena across different domains~\cite{SynchronizationComplexNetworks2014dorfler}.

In this context, the node state $x_v$ represents the phase $\theta_v$ of an oscillator, and the class-intrinsic biases $b_{z_v}$ correspond to class-intrinsic natural frequencies $\omega_{z_v}$.
A nonlinear sinusoidal coupling mechanism $f(\theta_v, \theta_u) = \sin(\theta_u - \theta_v)$ between connected nodes governs the evolution of nodes, which leads to the following dynamical system:
\begin{align}\label{equ:kuramoto_system}
\dot{\theta}_v = \omega_{z_v} + \sum_{u \in N(v)} W_{z_v,z_u} \sin(\theta_u - \theta_v).
\end{align}
If the phases synchronise, all oscillators rotate at the same collective mean-field frequency $\Omega$ (such that $\dot{\boldsymbol{\theta}}=\Omega\mathbf{1}$), so the system reaches a relative equilibrium.
In other words, the phase configuration is stationary if viewed in a frame rotating at $\Omega$ \cite{SynchronizationComplexNetworks2014dorfler}.
This is what allows us to recover the relative coupling strengths $W^*$ and relative effective frequencies $\tilde{\omega}_z^*$.

\begin{figure}
    \centering
    \input{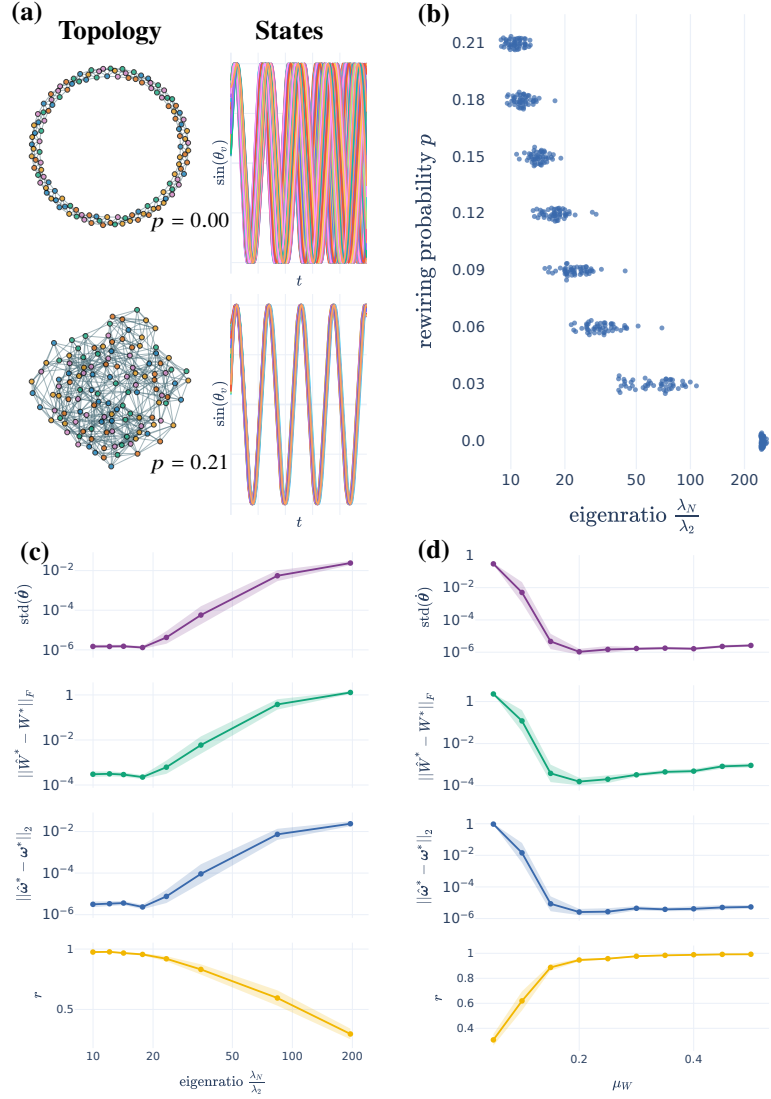}%
    \caption{\textbf{Experiments on nonlinear Kuramoto dynamics.} \textbf{(a)} Topologies and corresponding sine-transformed node states $\sin(\theta_v)$ across time for Watts-Strogatz random graphs with no ($p=0$) and high ($p=0.21$) rewiring probabilities. The comparison illustrates how small-world shortcuts foster rapid phase coherence. \textbf{(b)} The rewiring probability $p$ plotted against the corresponding eigenratio $\frac{\lambda_N}{\lambda_2}$ (log-scaled) of the graph Laplacian. Values are obtained from 50 Watts-Strogatz random graphs with $n=125$ nodes, $c=5$ uniformly distributed classes, and an initial connection to the $3$ nearest neighbours. \textbf{(c)} Velocity standard deviation $\operatorname{std}(\dot{\boldsymbol{\theta}})$, the normalised reconstruction error for the coupling strengths $\|\hat{W}^* - W^*\|_F$ and natural frequencies $\|\hat{\boldsymbol{\omega}}^* - \boldsymbol{\omega}^*\|_2$, and the Kuramoto order parameter $r$ evaluated at $t=100$ as a function of the eigenratio. The displayed averages and 95\% confidence intervals are binned and calculated identically to \Cref{fig:equilibrium_diffusion}. Coupling strengths and natural frequencies are drawn from uniform distributions: $\mu_W=\frac{1}{10}\pi$, $\sigma_W = \frac{1}{10^{2}\sqrt{3}}\pi$, $\mu_{\boldsymbol{\omega}}=\frac{5}{10}\pi$, and $\sigma_{\boldsymbol{\omega}} = \frac{2}{10\sqrt{3}}\pi$. \textbf{(d)} The same metrics evaluated across varying mean coupling strengths $\mu_W \in [0.05, 0.5]$ on Watts-Strogatz random graphs with a fixed rewiring probability of $p=0.21$, highlighting the phase transition to a locked state. The asterisk $*$ marks matrices and vectors as normalised (see Methods).}
    \label{fig:equilibrium_kuramoto}
\end{figure}

Similar to consensus dynamics, the topology influences the propensity of the system to reach a collective synchronised state, which we quantify using the eigenratio $\frac{\lambda_N}{\lambda_2}$ between the largest and the second smallest eigenvalues of the Laplacian matrix~\cite{MasterStabilityFunctions1998pecora,SynchronizationSmallWorldSystems2002barahona}.
We generate Watts-Strogatz random graphs with varying topologies (illustrated in \Cref{fig:equilibrium_kuramoto}a) and control this eigenratio using the rewiring probability $p$ in \Cref{fig:equilibrium_kuramoto}b.
\Cref{fig:equilibrium_kuramoto}c compares the velocity standard deviation and the reconstruction errors of our approach across these varying topological regimes.
Additionally, we quantify the macroscopic phase coherence of the network using the Kuramoto order parameter $r = \left\lvert\frac{1}{n}\sum_{v\in V} e^{i\theta_v}\right\rvert \in [0,1]$, where larger values indicate higher phase coherence~\cite{ChemicalOscillationsWaves1984kuramoto}.

As expected, our method performs well for systems that successfully reach a coherent, phase-locked state ($r \to 1$).
With decreasing order parameter $r$, the standard deviation of the velocities and the reconstruction error both increase, confirming that our method requires a phase-locked state where the system achieves uniform velocity.
In contrast to linear consensus dynamics, where a steady drift state is reached eventually in a connected graph, nonlinear Kuramoto dynamics are not guaranteed to synchronise.
The network undergoes a phase transition into a locked state only when the mean coupling strength $\mu_W$ is sufficiently large to overcome the dispersion of the natural frequencies~\cite{CriticalCouplingKuramoto2011dorfler}.
\Cref{fig:equilibrium_kuramoto}d visualises this transition by controlling the mean coupling strength $\mu_W$. 
Moving from left to right, the system undergoes a phase transition from an incoherent phase to a synchronised phase with a large order parameter ($r\to 1$).
Consistent with our theory, as the system crosses this critical coupling strength threshold, the standard deviation of the individual oscillator velocities converges to zero.
Precisely at this transition to uniform velocity, the reconstruction error decreases by several orders of magnitude, reaffirming that velocity uniformity is the prerequisite for accurate parameter inference.

\paragraph{Learning Absolute Coupling Strengths}
With the inference approach presented above, we recover the normalised coupling strengths and effective biases or frequencies.
This yields a relative understanding of the inter-class influences.
However, we can go a step further and infer the exact, absolute class-based coupling strengths and class-intrinsic bias values if we include an additional leakage term ($-x_v$), which -- in the absence of influence from a node's neighbours -- anchors the node state to the class-based bias.
This gives rise to the following dynamical system:
\begin{align}\label{equ:ordinary_dynamical_system}
    \dot{x}_v = b_{z_v} - x_v + \sum_{u \in N(v)} W_{z_v,z_u} f(x_v,x_u).
\end{align}
A consequence of the leakage term $-x_v$ in \Cref{equ:ordinary_dynamical_system} is that we can rewrite the dynamics as the following inhomogeneous system at the stationary state with $\dot{\mathbf{x}} = \mathbf{0}$:
\begin{align}
    x_v = b_{z_v} + \sum_{u \in N(v)} W_{z_v,z_u} f(x_v, x_u).
\end{align}
Unlike in the homogeneous formulation, the trivial zero-solution is eliminated, and the parameters can be recovered via Ordinary Least Squares (see Methods).

We illustrate our approach by assuming another coupling mechanism $f(x_v, x_u) = x_u$, which is motivated by a biological setting. In this context, the graph structure reflects the spatial connectivity of cells in a tissue environment, where different classes correspond to cell types. Node states represent cell type-specific gene programs, and coupling between neighbouring cells, e.g., due to molecular signalling or physical interactions, induces crosstalk of gene programs in spatial neighbours. As a first-order approximation, these couplings are assumed to be linear, but nonlinear couplings could be accommodated by our model, as shown above.

\begin{figure}
    \centering
    \input{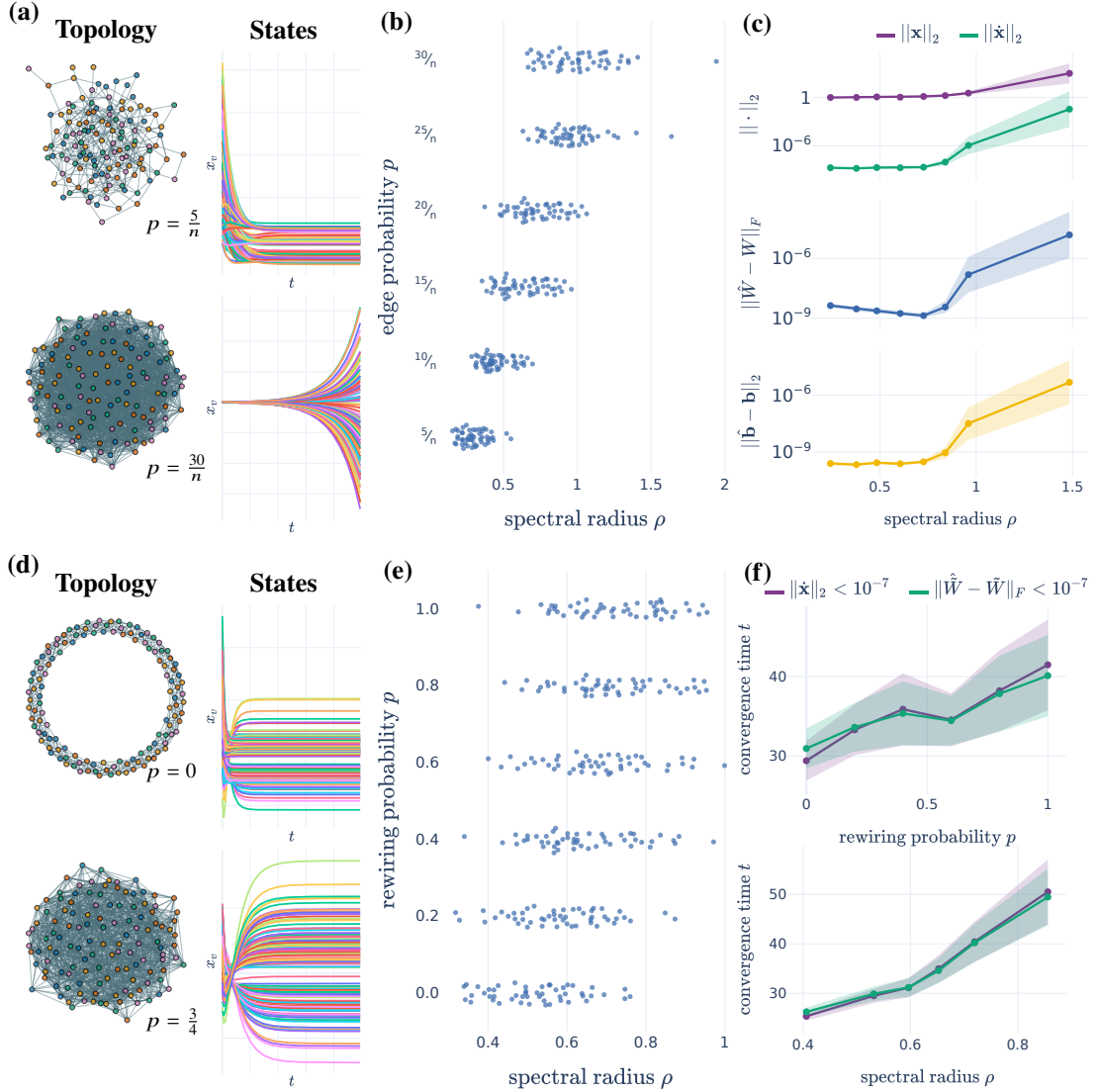}%
    \caption{\textbf{Experiments on anchored dynamics.} \textbf{(a)} Topologies and corresponding node state trajectories for Erd\H{o}s-R\'{e}nyi random graphs with sparse ($p=\frac{5}{n}$) and dense ($p=\frac{30}{n}$) connectivity. The sparse system successfully converges to a stable stationary state, while the dense system diverges. \textbf{(b)} The spectral radius $\rho$ (the largest absolute eigenvalue of the weighted adjacency matrix) as a function of the edge probability $p$ for Erd\H{o}s-R\'{e}nyi random graphs. \textbf{(c)} The node state and velocity norm ($\|\mathbf{x}\|_2$, $\|\dot{\mathbf{x}}\|_2$), and the \emph{absolute, unnormalised} reconstruction error of the weights ($\|\hat{W} - W\|_F$) and biases ($\|\hat{\mathbf{b}} - \mathbf{b}\|_2$) as a function of the spectral radius $\rho$. Systems in the stable regime ($\rho < 1$) converge to a stationary equilibrium, allowing near-perfect parameter recovery via Ordinary Least Squares. Systems crossing the instability threshold ($\rho \ge 1$) diverge towards infinity ($\|\mathbf{x}\|_2 \to \infty$), causing the inference to fail. \textbf{(d)} Topologies and corresponding node state trajectories for Watts-Strogatz random graphs with no ($p=0$) and high ($p=\frac{3}{4}$) rewiring probability, showing that the regular lattice network converges faster towards equilibrium than a heavily rewired topology. \textbf{(e)} The spectral radius $\rho$ as a function of the rewiring probability $p$ for Watts-Strogatz random graphs. Because the weighted adjacency matrix determines $\rho$, networks with identical structural rewiring $p$ exhibit a wide spread in $\rho$ driven by the randomly sampled weights $W$. \textbf{(f)} The required convergence time $t$ of the dynamical system ($\|\dot{\mathbf{x}}\|< 10^{-7}$) and the convergence time $t$ to achieve highly accurate parameter inference (measured via threshold $\|\hat{W} - W\|_F < 10^{-7}$). While the convergence time loosely correlates with the structural rewiring probability $p$ (top), it scales much more tightly with the actual spectral radius $\rho$ (bottom). System parameters are matched to \Cref{fig:equilibrium_diffusion}, with weights and biases sampled from $W_{z_v,z_u}\sim \mathcal{N}(0.0,0.075)$ and $b_{z_v}\sim \mathcal{N}(0.0,0.1)$.}
    \label{fig:equilibrium_ordinary}
\end{figure}

For the resulting dynamical system to reach a stationary state, the spectral radius $\rho$ (the largest absolute eigenvalue of the weighted adjacency matrix) must satisfy $\rho < 1$~\cite{MatrixAnalysis2012horn}. 
To investigate the accuracy of our method in relation to this condition, we generate Erd\H{o}s-R\'{e}nyi random graphs with varying edge probabilities $p$ (illustrated in \Cref{fig:equilibrium_ordinary}a).
The generated networks exhibit a wide range of spectral radii, as shown in \Cref{fig:equilibrium_ordinary}b.
To assess whether we are able to infer the \emph{absolute} class-based coupling strengths, we measure the reconstruction error \emph{without normalising the obtained parameters} and compare it to the norms of the node states and velocities in \Cref{fig:equilibrium_ordinary}c.
As expected, the plot shows vanishingly small velocity norms and near-perfect reconstruction errors in the stable regime ($\rho < 1$).
Conversely, the dynamics diverge as the system crosses the instability threshold ($\rho\geq1$).
This physical divergence is clearly visible as the norm of the node states $\|\textbf{x}\|_2$ explodes.
Because the system fails to reach a stationary state, the velocities never vanish, leading to a simultaneous surge in the reconstruction error by several orders of magnitude.

In the stable regime ($\rho < 1$), the spectral radius also governs the convergence rate at which the dynamical system reaches its stationary state; specifically, the system converges faster as $\rho$ approaches $0$.
We investigate this convergence speed and its relation to our inference method using Watts-Strogatz ring lattice graphs with varying rewiring probabilities $p$ (illustrated in \Cref{fig:equilibrium_ordinary}d). 
\Cref{fig:equilibrium_ordinary}e shows the spread of spectral radii for these different rewiring probabilities $p$.
We observe that while increased rewiring leads to higher average values for $\rho$, the massive spread of $\rho$ for a specific rewiring probability $p$ indicates that the specific, randomly sampled coupling strengths exert a stronger influence on the system than the structural rewiring itself.
This conclusion is supported by \Cref{fig:equilibrium_ordinary}f, which plots the time required for both the node velocities and the reconstruction error to converge.
While the convergence time shows only a loose, highly variant correlation with the rewiring probability $p$ (top), it scales more directly with the spectral radius $\rho$ (bottom).
Furthermore, the identical scaling behaviour between the convergence of the node velocities and the decay of the reconstruction error confirms our theoretical assumption that velocity uniformity dictates inference accuracy.

Interestingly, this dynamical system, driven by direct neighbour-state coupling ($f(x_v, x_u) = x_u$), converges faster for a more regular, lattice-like graph topology.
Mathematically, such faster convergence occurs because the convergence speed of this system is determined by the spectral radius of the adjacency matrix, which is constrained by the maximum and average degree of a network~\cite{RobustnessNetworksViruses2006jamakovica,MatrixAnalysis2012horn}. 

\section*{Discussion}
\label{sec:discussion}
Our results show that a single snapshot of node states can suffice to recover hidden class-based coupling strengths in network dynamics, without perturbing the system and without access to its temporal evolution.
The approach is indifferent to the coupling mechanism, provided the functional form is known.
It evaluates the coupling function on the observed states and freezes any coupling function into fixed coefficients, meaning the same inference applies to linear and nonlinear dynamics alike.
This is possible because a common velocity across nodes absorbs the collective motion into an effective bias, so that a single configuration constrains the couplings.
As in inverse statistical mechanics \cite{InverseStatisticalProblems2017nguyen}, rather than deriving a configuration from known interactions, our approach holds the observed configuration fixed and infers the interactions consistent with it.
We obtain the variation that time-series methods accumulate over successive instants instead from the diversity of local neighbourhoods.
Without a heterogeneous, class-specific bias, however, the system relaxes into a homogeneous configuration in which all pairwise differences vanish and the information about the coupling strengths is destroyed.
When the snapshot is taken at a true stationary state, an anchored variant of the formulation instead recovers the coupling matrix in absolute terms — the same problem solved by perturbation-based reverse engineering of gene regulatory networks \cite{PerturbSeqDissectingMolecular2016dixit}.
Our previous work extended such approaches to spatial regulatory networks of genes or gene programs across cells co-localised in tissues \cite{NiCoIdentifiesExtrinsic2024agrawal}.
Here, we provide the theoretical foundations and constraints for learning these interactions from the spatial heterogeneity of a single passive snapshot rather than from many controlled perturbations.

Our method further assumes known classes and a known coupling mechanism, requires a regime of uniform velocity that chaotic or unsynchronised dynamics might not reach, and, in its homogeneous formulation, recovers couplings only up to scale and shift.
These limitations naturally point towards future directions such as robustness to noise in the node states, network topology, or class assignments; inference without a known coupling mechanism; and extensions to higher-order couplings.
More broadly, our results indicate that spatial heterogeneity can serve as a substitute for temporal sampling, so that a single equilibrium configuration can carry sufficient information about network couplings to make repeated observation unnecessary.

\section*{Methods}
\label{sec:methods}
\paragraph{General Homogeneous System Formulation}
Our goal is to solve the inverse problem of reconstructing the class-wise coupling matrix $W \in \mathbb{R}^{c \times c}$ and the bias vector $\mathbf{b} \in \mathbb{R}^c$ from the network's adjacency matrix $A\in \left\{0,1\right\}^{n\times n}$ and a single snapshot of the node states $\mathbf{x}$.
To achieve this, we cast the dynamics into matrix notation:
\begin{align}\label{equ:general_dynamical_system_matrix}
    \dot{\mathbf{x}} = C \mathbf{b} + \left(CWC^T \odot A \odot S\right) \cdot \mathbf{1}
\end{align}
where $C \in \{0, 1\}^{n\times c}$ is the node class assignment matrix with $C_{v,z_v} = 1$ and 0 otherwise, $S \in \mathbb{R}^{n\times n}$ is the state-dependent coupling matrix with $S_{v,u}=f(x_v,x_u)$, $\odot$ denotes the Hadamard product, and $\mathbf{1}$ is a vector of ones.

We can infer these parameters from any instantaneous snapshot where the system exhibits a uniform velocity across all nodes -- meaning a relative equilibrium $\dot{\mathbf{x}} = k\mathbf{1}$ for some scalar $k$.
This relative equilibrium manifests, for example, as steady-state drift in linear consensus dynamics or as phase-locked synchrony in the Kuramoto model.
Our framework mathematically only requires this spatial velocity uniformity at the moment of measurement, which yields
\begin{align}\label{equ:general_dynamical_system_equilibrium}
    k\mathbf{1} = C \mathbf{b} + \left(CWC^T \odot A \odot S\right) \cdot \mathbf{1}.
\end{align}
In practice, a snapshot may only approximate this condition, so we need a way to quantify how far we are away from a relative equilibrium.
When we decompose the velocity into its mean and a zero-mean residual $\dot{\mathbf{x}} = \bar{v}\mathbf{1} + \delta\dot{\mathbf{x}}$, the mean $\bar{v}$ is the shared collective rate $k$ while the residual $\delta\dot{\mathbf{x}}$ is what must vanish at a relative equilibrium.
Since $\|\delta\dot{\mathbf{x}}\|_2 = \sqrt{n}\,\operatorname{std}(\dot{\mathbf{x}})$, the standard deviation measures exactly this residual.
So we can quantify the closeness to a relative equilibrium by the standard deviation of the node velocities $\operatorname{std}(\dot{\mathbf{x}})$.

By defining an effective bias $\tilde{b}_z = b_z-k$, we can move the constant $k$ to the right side and treat the problem as a homogeneous system.
For each class $z$, we define a design matrix based on the local neighbourhood interactions \cite{NetworkbasedNeighborhoodRegression2025zhen}:
\begin{align}\label{equ:M_z_general}
    M_z = \operatorname{diag}\left(C_{\cdot,z}\right)\left(S \odot A\right) C \in \mathbb{R}^{n\times c},
\end{align}
where $\operatorname{diag}(\cdot)$ creates a matrix from a vector and inserts the vector entries along the diagonal of the matrix.
The matrix $M_z$ represents the class-specific local neighbourhood structure for nodes of class $z$.
Each row corresponds to a node, and each column $\tilde z$ aggregates the coupling terms $f(x_v,x_u)$ contributed by neighbours belonging to class $\tilde z$.
Rows associated with nodes outside class $z$ are zero.
Thus, $M_z$ encodes, for every node of class $z$, the total coupling signal received from each neighbouring class (see \Cref{fig:M_z} for an illustration).

To incorporate the bias, we define the augmented design matrix $\tilde{M}_z = \left[ M_z | C_{\cdot,z}\right]$ and parameter vector $\tilde{W}_{z,\cdot} = \left[ W_{z, \cdot} | \tilde{b}_z \right]^T$.
For snapshots taken only close to relative equilibrium, the homogeneous equations are satisfied only approximately.
In this numerical setting, we require the augmented design matrix $\tilde{M}_z$ to have a one-dimensional null space, so that the total-least-squares solution is determined uniquely up to scale.
Provided the effective bias is non-zero $(\tilde{b}_z \not = 0)$, the appended column lies in the column space of $M_z$, so $\operatorname{r}(\tilde{M}_z) = \operatorname{r}(M_z)$.
This requirement is therefore equivalent to $M_z$ having full column rank $c$, which we check in practice.

By formulating the parameter inference as a homogeneous system, we recover the relative coupling strengths and effective biases, rather than their absolute numerical values.
The shift arises because we never observe the system's uniform velocity $k$.
Since $k$ enters the formulation only through the combination $\tilde{b}_z = b_z - k$, every inferred bias is displaced from its true value by this same unknown constant.
We can therefore resolve how the biases differ between classes, but not their absolute level.
The scale ambiguity arises from the form of the equations themselves.
Once the effective bias is moved to the right-hand side, each class obeys $\tilde{M}_z \tilde{W}_{z,\cdot} = \mathbf{0}$ -- a set of equations whose right-hand side is exactly zero. Such equations constrain only the ratios among the parameters, never their overall magnitude.
To see this, suppose some choice of parameters $\tilde{W}_{z,\cdot}$ satisfies $\tilde{M}_z \tilde{W}_{z,\cdot} = \mathbf{0}$.
Multiplying every parameter by a common constant $\alpha$ gives $\tilde{M}_z (\alpha \tilde{W}_{z,\cdot}) = \alpha\, \tilde{M}_z \tilde{W}_{z,\cdot} = \mathbf{0}$, which is satisfied equally well.
The observed states thus cannot distinguish a solution from any uniformly rescaled copy of it, and the coupling parameters are fixed only up to a common multiplicative factor.

By inserting the coupling mechanism $f(x_v,x_u)$ of a specific dynamical system, we can define the elements of the state-dependent coupling matrix and assemble the augmented design matrix $\tilde{M}_z$ for each class z of this system as explained above.
The solution $\hat{\tilde{W}}_{z,\cdot}$ is extracted via Singular Value Decomposition (SVD) as the right singular vector corresponding to the smallest singular value.
With this approach, we can solve the inference problem for both examples used above, i.e. linear consensus dynamics with $f(x_v,x_u) = x_u - x_v$ and Kuramoto dynamics with $f(x_v,x_u) = \sin(\theta_u - \theta_v)$.

The reason a single formulation accommodates both linear and nonlinear coupling mechanisms is that the inference is linear in the unknown parameters, not in the node states.
The functional form $f$ may depend on $x_v$ and $x_u$ in an arbitrarily nonlinear way, but at a fixed snapshot, the states are observed quantities.
Each entry $S_{v,u} = f(x_v, x_u)$ of the coupling matrix therefore evaluates to a fixed number, and so do all entries of the design matrix $\tilde{M}_z$, which is assembled from $S$ together with the known adjacency and class-assignment matrices.
The unknown coupling strengths and biases enter only through the vector $\tilde{W}_{z,\cdot}$ that this matrix multiplies.
Consequently, the governing relation $\tilde{M}_z \tilde{W}_{z,\cdot} = \mathbf{0}$ is a linear system in the unknowns, regardless of the form of $f$:
the nonlinearity is confined entirely to the numerical coefficients of $\tilde{M}_z$ and never reaches the parameters we solve for.
Evaluating $f$ at the snapshot, in other words, freezes the nonlinear coupling into constant coefficients and reduces a nominally nonlinear inverse problem to linear algebra.

\paragraph{Ordinary Least Squares for Anchored Dynamics}
The homogeneous formulation inherently yields only relative parameters, as it is subject to both an additive shift from absorbing the uniform velocity $k$ into the effective bias, and a scale ambiguity arising from the null-space problem.
To recover absolute coupling strengths, we instead consider a system where each node's evolution includes a leakage term $(-x_v)$, anchoring the dynamics to the absolute node state:
\begin{align}\label{equ:ordinary_dynamical_system_2}
    \dot{x}_v = b_{z_v} - x_v + \sum_{u \in N(v)} W_{z_v,z_u} f(x_v,x_u).
\end{align}
In contrast to our homogeneous formulation -- which requires only an instantaneous uniform velocity -- inferring the absolute parameters requires the system to reach a true temporal stationary state ($\dot{\mathbf{x}}=\mathbf{0}$).
Under this equilibrium condition, the temporal derivative vanishes, and we can rearrange the equation to isolate the known node state on the left-hand side:
\begin{align}
    x_v = b_{z_v} + \sum_{u \in N(v)} W_{z_v,z_u} f(x_v,x_u).
\end{align}
Because the target variable $x_v$ is now known and isolated, the trivial zero-solution is eliminated.
This transforms the parameter inference from a homogeneous problem into an inhomogeneous linear system that can be solved using Ordinary Least Squares (OLS) to recover the absolute coupling weights $W_{z_v,z_u}$ and biases $b_{z_v}$.

\paragraph{Reconstruction Error}
We quantify the accuracy of the inference as the discrepancy between the normalised estimated and ground-truth parameters defined above.
For the relative formulation, we report the reconstruction error of the class-based coupling strengths as the Frobenius norm of the difference between the normalised matrices, $\|\hat{W}^* - W^*\|_F$, and that of the effective biases as the Euclidean norm $\|\hat{\mathbf{b}}^* - \mathbf{b}^*\|_2$ (with the biases replaced by the class-intrinsic natural frequencies $\hat{\boldsymbol{\omega}}^*, \boldsymbol{\omega}^*$ in the Kuramoto case).
The Frobenius norm (and the Euclidean norm in the vector case) is the sum over the entrywise root of the squared deviations, so an error approaching zero signifies that every inter-class coupling has been recovered, up to the scale and shift freedoms removed by normalisation.
For the anchored dynamics, where the absolute parameters are recovered via Ordinary Least Squares, and no normalisation is required, we report the corresponding unnormalised errors $\|\hat{W} - W\|_F$ and $\|\hat{\mathbf{b}} - \mathbf{b}\|_2$ directly.
Unless stated otherwise, all reported errors are averaged across the random realisations of each experiment.

\paragraph{Parameter Normalisation}
To evaluate the reconstruction error, we compare the estimated parameters $\hat{W}$ and $\hat{\mathbf{b}}$ against the ground truth $W$ and $\mathbf{b}$.
As established above, our homogeneous formulation determines the parameters only up to a common multiplicative factor and, for the biases, an additional additive shift.
We therefore apply a normalisation procedure that undoes these two freedoms before comparison.
The common multiplicative factor carries an arbitrary sign, since multiplying every parameter of a class by $-1$ leaves the relation $\tilde{M}_z \tilde{W}_{z,\cdot} = \mathbf{0}$ satisfied.
We first fix this sign by multiplying each row of $\hat{W}$ by the sign of its entry of largest absolute value, thereby aligning the orientation of each estimated class-wise parameter vector with that of the ground truth.
We then remove the remaining magnitude by normalising the estimated and ground-truth coupling weights for each class $z$ row-wise to the range $[0, 1]$:
\begin{equation}
W_{z, \cdot}^* = \frac{W_{z, \cdot} - \min(W_{z, \cdot})}{\max(W_{z, \cdot}) - \min(W_{z, \cdot}) + \epsilon}
\end{equation}
where $\epsilon = 10^{-8}$ is a small constant for numerical stability.
The biases are subject to the same multiplicative factor and, in addition, to the unknown shift.
Because the uniform velocity $k$ is never observed, each true bias is masked by this constant through $\tilde{b}_z = b_z - k$;
we therefore compare against the effective ground-truth bias $\tilde{b}_z = b_z - k$ rather than $b_z$ itself.
To remove the multiplicative factor, both estimated and true biases are then divided by the largest absolute weight of their respective class and subsequently normalised to the range $[0, 1]$.
\section*{Acknowledgements}
This work was supported by the Federal Ministry of Research, Technology and Space in Germany, Grant No. 100582863 (TissueNet - 031L0311A).

\section*{Conflict of Interest}
D.G. serves on the scientific advisory board of Gordian Biotechnology. 

\section*{Author Contributions}

M.L., D.G.\ and I.S.\ conceived and planned the study.
M.L.\ performed all experiments and analyses and wrote the manuscript.
D.G.\ and I.S.\ supervised the research and contributed to writing and revising the manuscript.

\newpage
\bibliography{bibliography}

\newpage

\begin{edfigure}[bp]
    \centering
    \resizebox{\textwidth}{!}{%
\begin{tikzpicture}[
    >=Latex, font=\small,
    vertex/.style={
        circle, draw=black, minimum width=0.65cm, minimum height=0.65cm,
        font=\bfseries\fontsize{9}{9}\selectfont,
        fill opacity=0.75, draw opacity=0.75, text opacity=1, inner sep=0pt,
        text=black
    },
    class0/.style={fill=custom1}, class1/.style={fill=custom2}, class2/.style={fill=custom3},
    dimnode/.style={fill opacity=0.18, draw opacity=0.2, text opacity=0.25},
    rlabel/.style={circle, draw=black, minimum width=0.5cm, minimum height=0.5cm,
                   font=\bfseries\fontsize{7}{7}\selectfont, fill opacity=0.75, text opacity=1},
    qmark/.style={text=red, fill=white, inner sep=1pt, font=\bfseries\fontsize{8}{8}\selectfont, pos=0.5},
    myedge/.style={<-, >=Latex, dashed, bend left=15, opacity=0.85, line width=0.8pt, black},
    myedgedim/.style={<-, >=Latex, dashed, bend left=15, opacity=0.25, line width=0.6pt, black!50},
    mcell/.style={draw=black, minimum size=0.9cm, inner sep=0pt, font=\small},
    crossed/.style={mcell, text=custom4, font=\bfseries},
    agg/.style={->, line width=0.7pt, opacity=0.85, black!55},
    note/.style={font=\scriptsize, text=black!55, align=center},
]
\useasboundingbox (-3.75,-4.75) rectangle (10,4.75);


\node[vertex, class0] (A) at (0.1, 0.5) {A};
\node[vertex, class1] (B) at (-3.0, 1.0)    {B};
\node[vertex, class1, dimnode] (C) at (-0.7, -2.4)   {C};
\node[vertex, class0] (D) at (1.0, 1.8)   {D};
\node[vertex, class2] (E) at (0.2, -1.0) {E};
\node[vertex, class0] (F) at (-1.0, 2.0)    {F};
\node[vertex, class0] (G) at (-2.0, -1.0) {G};
\node[vertex, class2] (H) at (-1.5, 0.5)  {H};
\node[vertex, class1] (I) at (-2.3, -2.6)  {I};
\node[vertex, class2, dimnode] (J) at (-0.1, 3.0)  {J};

\draw[myedge] (A) to node[qmark] {?} (D);
\draw[myedge] (A) to node[qmark] {?} (E);
\draw[myedge] (A) to node[qmark] {?} (F);
\draw[myedge] (A) to node[qmark] {?} (G);
\draw[myedgedim] (B) to (F);
\draw[myedgedim] (C) to (E);
\draw[myedgedim] (C) to (G);
\draw[myedgedim] (F) to (A);
\draw[myedge] (G) to node[qmark] {?} (B);
\draw[myedgedim] (H) to (B);
\draw[myedgedim] (H) to (E);
\draw[myedgedim] (H) to (F);
\draw[myedgedim] (I) to (C);
\draw[myedge] (G) to node[qmark] {?} (I);
\draw[myedgedim] (J) to (F);
\draw[myedgedim] (D) to (J);
\draw[myedge] (G) to node[qmark] {?} (H);
\draw[myedgedim] (E) to (G);
\draw[myedgedim, bend left=-15] (J) to (B);

\draw[draw=custom1, thick, dash pattern=on 3pt off 2pt, smooth cycle, tension=0.7]
    plot coordinates {(-2.45,-1.15) (0.41,-1.36) (0.70,0.55) (1.37,2.20) (-1.24,2.30) (-1.1,0.4)};
\draw[draw=custom1, thick, densely dashed, smooth cycle, tension=0.5]
    plot coordinates {(-2.65,-2.93) (-1.69,-2.57) (-1.05,0.73) (-3.44,1.19)};
\node[note, anchor=south west, text=custom1, font=\small] at (-2.05,2.25) {$N(A)$};
\node[note, anchor=north east, text=custom1, font=\small] at (-1.9,-3.2) {$N(G)$};


\node[mcell, fill=custom1!57, text=white] (a1) at (3.3,3.0) {$-0.15$};
\node[mcell, fill=custom2!20]             (a2) at (4.2,3.0) {$0.00$};
\node[mcell, fill=custom3!89, text=white] (a3) at (5.1,3.0) {$0.28$};
\node[mcell, dashed]                      (ab) at (6.2,3.0) {$1$};
\node[rlabel, class0] at (2.3,3.0) {$A$};
\draw[line width=1.1pt, rounded corners=2pt] (2.85,2.55) rectangle (6.65,3.45);

\node[crossed] at (3.3,2.1){$\times$}; \node[crossed] at (4.2,2.1){$\times$};
\node[crossed] at (5.1,2.1){$\times$}; \node[crossed, dashed] at (6.2,2.1){$\times$};
\node[rlabel, class1] at (2.3,2.1) {$B$};
\node[crossed] at (3.3,1.2){$\times$}; \node[crossed] at (4.2,1.2){$\times$};
\node[crossed] at (5.1,1.2){$\times$}; \node[crossed, dashed] at (6.2,1.2){$\times$};
\node[rlabel, class1] at (2.3,1.2) {$C$};

\node[mcell, fill=custom1!20]             at (3.3,0.3) {$0.00$};
\node[mcell, fill=custom2!20]             at (4.2,0.3) {$0.00$};
\node[mcell, fill=custom3!56, text=white] at (5.1,0.3) {$0.15$};
\node[mcell, dashed]                      at (6.2,0.3) {$1$};
\node[rlabel, class0] at (2.3,0.3) {$D$};

\node[crossed] at (3.3,-0.6){$\times$}; \node[crossed] at (4.2,-0.6){$\times$};
\node[crossed] at (5.1,-0.6){$\times$}; \node[crossed, dashed] at (6.2,-0.6){$\times$};
\node[rlabel, class2] at (2.3,-0.6) {$E$};

\node[mcell, fill=custom1!60, text=white] at (3.3,-1.5) {$0.16$};
\node[mcell, fill=custom2!20]             at (4.2,-1.5) {$0.00$};
\node[mcell, fill=custom3!20]             at (5.1,-1.5) {$0.00$};
\node[mcell, dashed]                      at (6.2,-1.5) {$1$};
\node[rlabel, class0] at (2.3,-1.5) {$F$};

\node[mcell, fill=custom1!20]              (g1) at (3.3,-2.4) {$0.00$};
\node[mcell, fill=custom2!100, text=white] (g2) at (4.2,-2.4) {$0.33$};
\node[mcell, fill=custom3!25]              (g3) at (5.1,-2.4) {$-0.02$};
\node[mcell, dashed]                       (gb) at (6.2,-2.4) {$1$};
\node[rlabel, class0] at (2.3,-2.4) {$G$};
\draw[line width=1.1pt, rounded corners=2pt] (2.85,-2.85) rectangle (6.65,-1.95);

\node[font=\Huge] at (4.75,-3.3) {$\vdots$};
\node[note, anchor=west, text width=1.9cm, align=left] at (6.9,2.2) {$B,C,E,\dots$\\ not blue classes $\Rightarrow$ zeroed out};

\draw[line width=1pt] (2.85,3.45) -- (2.7,3.45) -- (2.7,-3.6) -- (2.85,-3.6);
\draw[line width=1pt] (6.65,3.45) -- (6.8,3.45) -- (6.8,-3.6) -- (6.65,-3.6);

\node[font=\Huge] at (7.15,-0.0) {$\cdot$};
\node[font=\Large] at (8.0,-0.075) {$
     \underbrace{\begin{bmatrix}\textcolor{custom1}{W_{1,1}}\\[2pt] \textcolor{custom2}{W_{1,2}}\\[2pt]
         \textcolor{custom3}{W_{1,3}}\\[2pt] \tilde b_1\end{bmatrix}}_{\textcolor{red}{\text{\Large\bfseries ?}}}
$};
\node[font=\Huge] at (9.40,-0.0) {$=\mathbf{0}$};

\node[note, anchor=north, text width=4.5cm, align=center] at (2.7,-4.2)
   {$\sum_{w \in N_z(G)} f(x_G, x_{w})$ per neighbour class $z$};

\draw[agg] (D) to[out=60,in=120]   ($(a1.north)!0.5!(a1.north west)$);
\draw[agg] (F) to[out=15,in=90]   (a1.north);
\draw[agg] (G) to[out=115,in=60,looseness=1.4]  ($(a1.north)!0.5!(a1.north east)$);
\draw[agg] (E) to[out=135,in=90,looseness=1.25]  (a3.north);
\draw[agg] (B) to[out=-100,in=-90] (g2.south);
\draw[agg] (I) to[out=-15,in=-120]  ($(g2.south)!0.5!(g2.south west)$);
\draw[agg] (H) to[out=-80,in=-90] (g3.south);

\end{tikzpicture}%
    }
    \caption{\textbf{Construction of the class-based design matrix $M_1$.} Illustrated on the same 10-node Kuramoto network and snapshot as \Cref{fig:overview}. Left: two selected class-1 nodes $A$ and $G$ (blue), together with their neighbourhoods $N(A)$ and $N(G)$ (dashed outlines). Right: each class-1 node contributes one row of $M_1$; arrows trace how each neighbour's known coupling value $f(x_v,x_u)$, evaluated at the snapshot, is summed by class into the corresponding cell (colour intensity shows the magnitude). Rows of nodes outside class 1 are zeroed by $\operatorname{diag}(C_{\cdot,1})$ and omitted after row $G$ for brevity. Augmented with a column of ones and multiplied by the unknown, augmented weight vector $\tilde W_{1,\cdot}$, every row of the resulting homogeneous system $\tilde{M}_1\cdot \tilde{W}_{1,\cdot} = \mathbf{0}$ holds at the relative equilibrium.}
    \label{fig:M_z}
\end{edfigure}

\end{document}